\documentclass[aps,prd,preprint,tightenlines,superscriptaddress,nofootinbib,showpacs]{revtex4}
\usepackage{amssymb,latexsym}
\usepackage{amsmath,amsbsy}
\usepackage{epsfig,bm}
\usepackage{graphicx,comment}
\DeclareMathOperator{\tr}{tr}

\def\a{{\alpha}}
\def\b{{\beta}}
\def\d{{\delta}}
\def\D{{\Delta}}
\def\e{{\varepsilon}}
\def\g{{\gamma}}

\def\L{{\Lambda}}

\def\O{{\Omega}}
\def\S{{\Sigma}}
\def\s{{\sigma}}

\def\X{{\Xi}}

\def\ol#1{{\overline{#1}}}

\def\CPT{{$\chi$PT}}

\def\cF{{\mathcal F}}

\def\cL{{\mathcal L}}
\def\cO{{\mathcal O}}
\def\cA{{\mathcal A}}

\def\cH{{\mathcal H}}

\def\cM{{\mathcal M}}
\def\cD{{\mathcal D}}

\def\cJ{{\mathcal{J}}}

\begin{document}




\preprint{UMD-40762-416}

\title{Hyperons in Two Flavor Chiral Perturbation Theory}

\author{Brian~C.~Tiburzi}
\email[]{bctiburz@umd.edu}
\affiliation{%
Maryland Center for Fundamental Physics, 
Department of Physics, 
University of Maryland, 
College Park,  
MD 20742-4111, 
USA
}

\author{Andr\'{e}~Walker-Loud}
\email[]{walkloud@wm.edu}
\affiliation{%
Maryland Center for Fundamental Physics, 
Department of Physics, 
University of Maryland, 
College Park,  
MD 20742-4111, 
USA
}
\affiliation{%
Department of Physics,
College of William and Mary,
Williamsburg, VA 23187-8795,
USA
}

\begin{abstract}
We use two-flavor  chiral perturbation theory to describe hyperons.  We focus on the strangeness conserving sector, and, as an example, calculate hyperon masses.  Convergence of this two-flavor chiral expansion for observables is improved over the three-flavor theory.  The cost, however, is a larger number of low-energy constants that must be ultimately determined from lattice QCD data. 
A formula for the mass of the omega baryon is derived to sixth order in this expansion, 
and will aid lattice practitioners in scale setting or tuning the strange quark mass.
\end{abstract}

\date{\today}


\maketitle



\section{Introduction}


The study of QCD at low energies is a challenging endeavor. 
In the low-energy spectrum,
the axial symmetry of the massless QCD Langrangian is hidden, 
and we identify the pseudoscalar
mesons with the (approximate) Goldstone 
bosons that emerge from spontaneous 
chiral symmetry breaking.  
This fact  can be incorporated into a phenomenological
effective field theory of low-energy QCD, 
called chiral perturbation theory 
(\CPT)~\cite{Weinberg:1978kz,Gasser:1983yg,Gasser:1984gg}. 
\CPT\ provides an effective description of low-energy QCD 
only if there is a systematic power counting to
order the infinite tower of operators needed
for renormalization.
For the Goldstone bosons of QCD, 
the small expansion parameter, 
$\e$, 
is 
$\e \sim p / \L_\chi$, 
where
$p$ is a typical meson momentum,
and 
$\L_\chi \approx 1 \, \texttt{GeV}$ 
is the chiral symmetry breaking scale.
The quark masses explicitly break chiral symmetry
giving rise to meson masses,
and so we must further treat
$m_{GB} / \L_\chi$ 
in the power counting. 
Looking at the masses of the lowest lying 
non-singlet pseudoscalar mesons, 
we find a typical expansion parameter
$\e \sim 0.1$ 
for a two-flavor chiral theory of just pions, 
and 
$\e \sim 0.5$ for a theory of the octet of mesons. 
Indeed it is questionable whether the 
strange quark mass is light enough 
to develop a consistent expansion about
the $SU(3)$ chiral limit. 
While meson observables have expansions in 
$\e^2$,
 baryons observables are complicated
by the nonvanishing chiral limit baryon mass 
$M_B$. 
Expanding about the nonrelativistic limit for baryons, 
one further encounters a series of terms that scale as 
$m_{GB} / M_B \sim \e$~\cite{Jenkins:1990jv,Jenkins:1991es}.

Two-flavor chiral expansions should exhibit better convergence over three-flavor expansions.  
Two-flavor \CPT\ has long been utilized for pions and nucleons. 
Calculations for strange hadrons have largely assumed $SU(3)$ chiral symmetry with a few exceptions.  For hyperons, there is a wealth of literature exploring hyperon properties utilizing the $SU(3)$ heavy baryon Lagrangian~\cite{Jenkins:1990jv,Jenkins:1991es,Jenkins:1991ts,Jenkins:1991bt,Jenkins:1991bs,Butler:1992ci,Jenkins:1992ab, Butler:1992pn,Jenkins:1992pi,Anderson:1993as, Butler:1993ht,Bernard:1993nj,Butler:1993ej,Lebed:1993yu,Lebed:1994gt,Lee:1994my,Savage:1994pf,Banerjee:1994bk,Savage:1995ca,Springer:1995nz,Bedaque:1995wb,Bedaque:1995pa,Banerjee:1995wz,Borasoy:1996bx,Savage:1996zd,Meissner:1997hn,Savage:1997me,AbdElHady:1998gw,Tandean:1998ch,Borasoy:1998pe,Borasoy:1999md,AbdElHady:1999mj,Kubis:1999xb,Kubis:2000aa,Borasoy:2000pq,Oller:2000fj,Frink:2004ic,Detmold:2005pt,Frink:2005ru,Villadoro:2006nj,Liu:2006xja,Lacour:2007wm,Jiang:2008aqa}.
On the other hand, an $SU(2)$ treatment of kaons was developed~\cite{Roessl:1999iu} to describe $\pi$--$K$ scattering near threshold, which was further explored in~\cite{Frink:2002ht}.  Strange matrix elements of the nucleon were explored with an $SU(2)$
Lagrangian~\cite{Chen:2002bz}, and an $SU(2)$ treatment of spin-$1/2$ baryons was pursued in~\cite{Beane:2003yx} to investigate hyper-nuclear interactions and hyperon decays at tree-level.  
In this work, we treat both spin-$1/2$ and spin-$3/2$ strange baryons using an $SU(2)$ Lagrangian.  
We utilize the two-flavor chiral expansion to calculate hyperon masses.  
Our study is motivated additionally by lattice QCD data.  

Lattice gauge theory~\cite{DeGrand:2006aa} is beginning to yield quantitative information about nonperturbative QCD.  Currently one must utilize \CPT\ to extrapolate lattice data to the physical quark masses.  As such, the convergence of \CPT\ expansions is relevant to control lattice systematics.  
Recent lattice studies suggest that $SU(3)$ chiral perturbation theory is insufficient to extrapolate lattice results for hyperons to the physical point~\cite{Lin:2007ap,WalkerLoud:2008bp} 
(also seen for some observables in the meson sector~\cite{Allton:2008pn,Aoki:2008sm,Mawhinney:Latt08}).  
Because the expansion in the baryon sector is expected to converge less rapidly than the meson sector, it is efficacious to treat hyperons is $SU(2)$. 
In the near future, however, lattice calculations will be performed at or near the physical 
point~\cite{Latt2008}.  
The role of $\chi$PT in conjunction with lattice QCD will then focus on reliably determining the unconstrained coefficients of the tower of operators in the chiral Lagrangian, known as the low energy constants (LECs).  These LECs can then be used to make predictions of physical processes which are difficult or impossible to directly access with experiment or lattice computational methods. 
The reliability and precision of these predictions will ultimately depend upon the convergence of the effective theory.
One practical lattice application of our work concerns the use of the $\O$ mass to either tune the strange quark
mass~\cite{Toussaint:2004cj} or set the scale~\cite{Lin:2007pt,Allton:2008pn} .  We derive an expression for the pion mass dependence of the $\O$ mass which allows a controlled extrapolation to the physical pion mass.

%
\section{$SU(2)$ Lagrangian for Hyperons}

%
\subsection{Strangeness $S=1$ Baryons \label{s:S=1L}}

The strangeness $S = 1$ baryons are of two varieties.  There are the spin-$1/2$ particles, the
$\L$ and the $\S$, as well as the spin-$3/2$ states, the $\S^*$.  The $\L$ is an isospin singlet, while both the $\S$ and $\S^*$ are isotriplets. It is convenient to package the $\S$ as
\begin{equation}
\S
= 
\begin{pmatrix}
\frac{1}{\sqrt{2}} \S^0  &
\S^+ \\
\S^- &
- \frac{1}{\sqrt{2}} \S^0
\end{pmatrix}
,\end{equation}
which transforms under the adjoint.
For the $\S^*$, we use a symmetric tensor
representation $\S^*_{ij}$, 
where:
$\S^*_{11} = \Sigma^{*+}$, 
$\S^*_{12} = \S^*_{21} = \frac{1}{\sqrt{2}} \S^{*0}$,
and
$\S^*_{22} = \S^{*-}$.
To have a consistent power counting, 
we remove $M_\L^{(0)}$, the chiral limit mass of the 
$\L$, 
using a field redefinition customary
for treating baryons as heavy particles~\cite{Jenkins:1990jv,Jenkins:1991es}. 
The Lagrangian for the 
$S = 1$ 
baryons 
can be written down as a series of terms
that scale with powers of the small 
parameter 
$\e$, where
$\e \sim k \sim m_\pi$, 
where 
$k$
is a typical residual momentum. 
To 
$\cO(\e^2)$
in this scheme,
we have the leading Lagrangian
\begin{eqnarray} \label{eq:LS=1}
\cL^{(S=1)}_2 &=&
	\ol \L \left( i v \cdot \partial \right) \L
		+ \tr \left[ \ol \S \left( iv \cdot \cD - \D_{\L \S} \right) \S \right]
			- \Big( \ol \S {}^{*\mu} \left[ i v \cdot \cD - \D_{\L \S^*} \right] \S^*_\mu \Big)
\notag \\ && 
	-\frac{\s_\L}{(4\pi f)}  \ol \L \, \L \, \tr \left( \cM_+ \right) 
	- \frac{\s_\S}{(4\pi f)}  \tr \left( \ol \S \, \S \right) \tr \left( \cM_+ \right)
	+ \frac{\ol \s_\S}{(4\pi f)}  \Big( \ol \S {}^{*\mu} \, \S^*_\mu \Big) \tr \left( \cM_+ \right) ,
\end{eqnarray}
where the isospin symmetric mass spurion can be written as 
$\cM_+ = \frac{1}{4}  m_\pi^2 ( U + U^\dagger)$ by making use of the Gell-Mann--Oakes--Renner relation for the pion and quark masses, which at leading order is 
$m_\pi^2 = 2B m_q $~\cite{GellMann:1968rz}.
Additional terms involving the mass spurion
must be written down when one works away from 
the isospin limit, however we work in the isospin limit throughout.
Appearing in the mass spurion is $U$, the exponential
of the pion fields,
$U = \xi^2 = \exp ( 2 i \phi / f)$,
where 
\begin{equation}
\phi = 
\begin{pmatrix}
\frac{1}{\sqrt{2}} \pi^0  & \pi^+ \\
\pi^- & - \frac{1}{\sqrt{2}} \pi^0
\end{pmatrix}
,\end{equation}
and in our conventions $f  \approx 130 \, \texttt{MeV}$. 
In Eq.~\eqref{eq:LS=1}, 
the parameter 
$\D_{\L\S}=M_\S^{(0)} - M_\L^{(0)}$ ($\D_{\L\S^*} = M_{\S^*}^{(0)} - M_\L^{(0)}$)
is the mass splitting between the 
$\S$  ($\S^*$)
and 
$\L$ 
in the chiral limit. 
Phenomenologically, we know the splitting at the physical pion mass is 
$\D_{\L \S}  \sim 77 \, \texttt{MeV}$ ($\D_{\L \S^*}  \sim 270 \, \texttt{MeV}$).
We will treat 
$\D_{\L\S} \sim \D_{\L\S^*} \sim m_\pi$ 
in our power counting.
The chirally covariant derivatives $\cD_\mu$ appearing above act as follows
\begin{eqnarray}
\cD_\mu \Sigma &=&  \partial_\mu \S + [ V_\mu, \Sigma],   
\quad
\text{and}
\quad
\left( \cD_\mu \Sigma^*_\nu \right)_{ij}
=
\partial_\mu (\Sigma^*_\nu)_{ij}
+ 
(V_{\mu})_{i} {}^k  (\Sigma^*_\nu)_{kj}
+
(V_\mu)_{j} {}^k  (\Sigma^*_\nu)_{ik}
,\end{eqnarray}
with $V_\mu$ as the vector field of pions, 
$V_\mu = \frac{1}{2} \left( \xi \partial _\mu \xi^\dagger + \xi^\dagger \partial_\mu \xi \right)$.
The interactions between 
the $S=1$ baryons are contained
in the $\cO(\e)$ interaction Lagrangian
\begin{eqnarray}
\cL^{(S=1)}
&=&
g_{\pi\S\S} 
\tr \left(
\ol \S  S^\mu \left[A_\mu , \S \right] 
\right)
+
2 g_{\pi\S^*\S^*}
\left(
\ol \S {}^{*\mu}  S \cdot A   \S^*_\mu   
\right)
+
g_{\pi\S^*\S}
\left(
\ol \S {}^{*\mu}  A_\mu  \S  
+
\ol \S   A^\mu   \S^*_\mu
\right)
\notag \\
&+&
\sqrt{\frac{2}{3}} g_{\pi\L \S}
\Big[
\tr \left( 
\ol \S \, S \cdot A 
\right) \L
+
\ol \L 
\tr \left(
S \cdot A  \S
\right)
\Big]
+ 
g_{\pi\L \S^*}
\Big[
\left( 
\ol \S {}^{*\mu}   A_\mu 
\right) \L
+
\ol \L 
\left(
A^\mu  \S^*_\mu
\right)
\Big]
\label{eq:LAxialS=1}
.\end{eqnarray}
In this interaction Lagrangian 
appears the axial-vector field of pions $A_\mu$, 
given by 
$A_\mu = \frac{i}{2} \left(  \xi \partial _\mu \xi^\dagger -  \xi^\dagger \partial_\mu \xi \right)$.  
The tensor products have been denoted with parentheses, and are defined as follows:
$\Big( \ol \S {}^* \cA \, \S^* \Big)
= 
\ol \S {}^{*ij}  \cA_{j} {}^k \, \S^*_{ki}$,
$\Big( \ol \S {}^* \cA \, \S \Big)
= 
\ol \S {}^{*ij}  \cA_{j} {}^k \, \S_{k} {}^l \, \epsilon_{li}$,
and
$\Big( \ol \S {}^* \cA \Big)
= 
\ol \S {}^{*ij}  \cA_{j} {}^k \, \epsilon_{ki}$.

%
\subsection{Strangeness $S=2$ Baryons \label{s:S=2L}}

The strangeness $S=2$ baryons are the cascades, which both
can be packaged as the spinors
$\X = 
(\X^0 , \X^- )^T$,
and
$\X^*_\mu
=
( \X^{* 0}_\mu , \X^{* -}_\mu )^T$
that transform as isodoublets under 
$SU(2)$. 
In the 
$S=2$ 
sector, we redefine the fields
to phase away 
$M_\Xi^{(0)}$, 
which is the 
$\Xi$ 
mass in the chiral limit. 
The free Lagrangian for the cascades can then 
written to 
$\cO(\e^2)$ 
as
\begin{eqnarray}
\cL^{(S =2)}_2 &=&
	\left( \ol \X \left[ i v \cdot \cD \right] \X \right)
	- \frac{\s_\X}{(4\pi f)} \, \left( \ol \X \, \X  \right) \, \tr \left( \cM_+ \right)
\notag \\ && 
	- \left( \ol \X {}^{*\mu} \left[ iv \cdot \cD  - \D_{\X \X^*} \right] \X^*_\mu \right)
	+ \frac{\ol \s_\X}{(4\pi f)} \, \left( \ol \X {}^{*\mu} \, \X^*_\mu \right) \, \tr \left( \cM_+ \right).
\label{eq:LS=2}
\end{eqnarray}
The chirally covariant derivative 
$\cD_\mu$
acts on both doublets in the same manner, 
namely
$(\cD_\mu \Xi )_i 
= 
\partial_\mu \Xi_i
+ 
(V_\mu)_{i} {}^{j} \Xi_j$.
The parameter 
$\D_{\X \X^*} = M_{\Xi^*}^{(0)} - M_{\Xi}^{(0)}$ 
appearing in the free Lagrangian
is the mass splitting between the 
$\X^*$ 
and
$\X$
in the chiral limit. 
Phenomenologically, we know at the physical pion mass, 
the splitting 
$\D_\X \sim 215 \, \texttt{MeV}$,
and hence we treat 
$\D_{\X \X^*} \sim m_\pi$ 
in our power counting. 
The leading interactions between the $S=2$ baryons
are contained in the $\cO(\e)$ interaction Lagrangian
\begin{equation}\label{eq:LAxialS=2}
\cL^{(S =2)}
=
2 g_{\pi\X\X} \, 
\left(
\ol \X  \, 
S \cdot A \,
\X
\right)
+ 
2 g_{\pi\X^* \X^*}
\left(
\ol \X {}^{*\mu} \, 
S \cdot A \, 
\X^*_\mu
\right)
+
g_{\pi\X^* \X}
\left[
\left(
\ol \X {}^{*\mu} \,
A_\mu \,
\X
\right)
+ 
\left(
\ol \X \,
A^\mu \,
\X^*_\mu
\right)
\right]
.\end{equation}

%
\subsection{Strangeness $S=3$ Baryons \label{s:S=3L}}

In this sector lives the lone 
$\O^-$ 
which transforms as an isoscalar under 
$SU(2)$. 
We describe this state with a heavy baryon field 
$\O_\mu$,
having phased away the chiral limit mass, 
$M_\O^{(0)}$,
using a field redefinition. 
The free Lagrangian for the 
$\O^-$ 
to
$\cO(\varepsilon^6)$
is written as
\begin{eqnarray}
\cL^{(S = 3)} &=&
	- \ol \O^\mu  \left( iv \cdot  \partial \right)  \O_\mu
	+\frac{\ol \sigma_\O}{(4\pi f)} \, \ol \O^\mu \O_\mu \, \tr  \left( \cM_+ \right)
	  - \frac{t_{\O}^A}{(4\pi f)}  \ol \O^\mu \O_\mu \tr \left( A^2 \right)
		\notag  \\ &&
+ \frac{t_{\O}^{M}}{(4\pi f)^3} \, \ol \O^\mu \O_\mu \, [\tr \left( \cM_+ \right) ]^2
	- \frac{T_{\O}^{AM}}{(4\pi f)^3}  \ol \O^\mu \O_\mu  \tr \left( A^2 \right) \tr \left( \cM_+ \right) 
	+ \frac{T_{\O}^{M}}{(4\pi f)^5}  \ol \O^\mu \O_\mu  [\tr \left( \cM_+ \right) ]^3
\label{eq:S=3L}
.\notag\\
\end{eqnarray}
Because the $\O^-$
is a singlet, the partial derivative
is automatically chirally
covariant.
Unlike the other hyperons, the 
$\O^-$
does not interact with other baryons. 
Baryon interactions arise only when
considering the $\D S \neq 0$ Lagrangian.
Calculations can consequently be performed to very 
high orders in the strangeness conserving sector with ease. 
For this reason, we have included the necessary
higher-order terms in the Lagrangian.%
\footnote{
We have omitted operators of the form $\ol \O_\mu \O^\mu \tr [( v \cdot A)^2]$
because their contribution to the omega mass cannot be distinguished from
operators involving $\tr (A^2)$. The same is not true, however, for all observables. 
}
 As we shall explicitly demonstrate in Sec.~\ref{s:S=3M}, there is an improved expansion for the 
 $\O^-$ mass as compared to other heavy baryons.

%
\section{$SU(2)$ Mass formula for Hyperons}

\begin{figure}
\epsfig{file=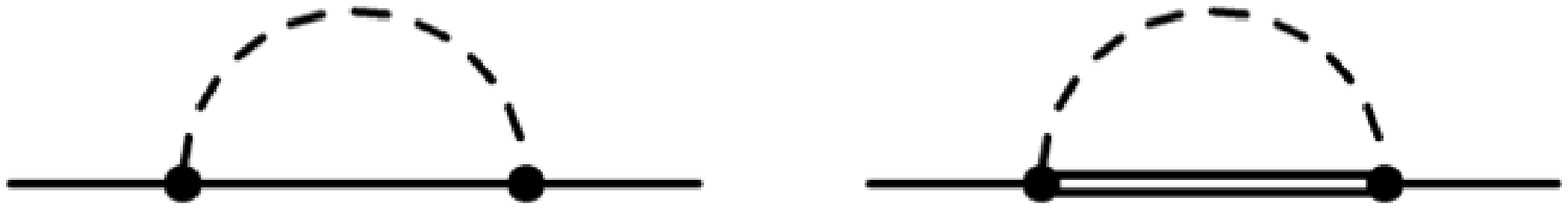,angle=270,width=6.7cm}
$\qquad \qquad$
\epsfig{file=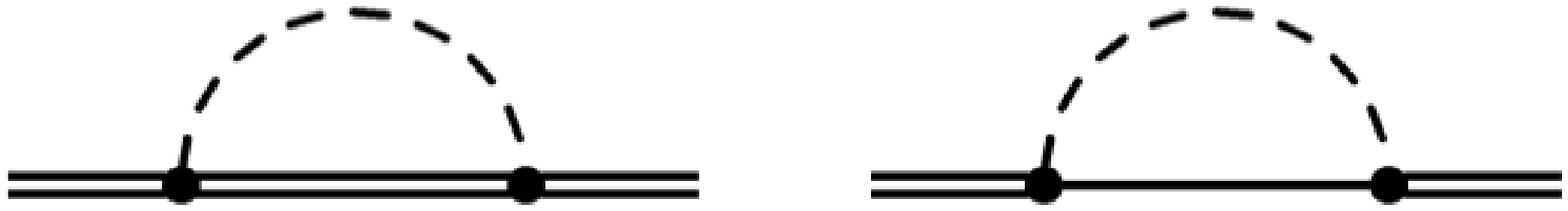,angle=270,width=6.7cm}
\caption{\label{f:mass} 
Loop diagrams for strangeness $S=1$, and $S=2$  baryon masses.
On the left (right) are contributing diagrams for spin-$1/2$ (spin-$3/2$) baryons. 
The filled circles denote axial couplings generated from the $\mathcal{O}(\e)$ interaction Lagrangian, 
Eqs.~\eqref{eq:LAxialS=1} and \eqref{eq:LAxialS=2}.}
\end{figure}

Having constructed the $SU(2)$ chiral Lagrangian for the strangeness
$S=1$, $2$, and $3$ 
sectors, we can calculate observables of interest.  Here we derive expressions for the pion mass dependence of the various hyperon masses. 
The hyperon masses all have an expansion of the form
\begin{equation}
M_B = M_B^{(0)} + c_B^{(2)} \frac{m_\pi^2}{(4\pi f_\pi)}
	+c_B^{(3)} \frac{m_\pi^3}{(4\pi f_\pi)^2}
	+c_B^{(4)} \frac{m_\pi^4}{(4\pi f_\pi)^3}
	+\dots\, ,
\end{equation}
modulo chiral nonanalytic functions. 
Here we use $c_B^{(n)}$ to denote a generic dimensionless coefficient 
of the term at $\cO(\e^n)$ in the expansion of the hyperon
mass about the chiral limit.  
A term of this order is always accompanied 
by a power of $(4 \pi f_\pi)^{1-n}$. 
The expansion starts out at 
$\mathcal{O}(\e^2)$, 
where there are tree-level insertions of the mass spurions given in Eqs.~\eqref{eq:LS=1}, \eqref{eq:LS=2}, and \eqref{eq:S=3L}.  
At third order, $\cO(\e^3)$, 
there are one-loop graphs generated from the various axial couplings in 
Eqs.~\eqref{eq:LAxialS=1} and \eqref{eq:LAxialS=2}.  
These one loop graphs are depicted in Fig.~\ref{f:mass}.  
%
%
%
%
These topologies are not present for the $\O^-$ baryon,
and we discuss  $M_\O$ separately in Sect.~\ref{s:S=3M}. 
Beyond this order, we have terms at $\cO(\e^4)$. 
Although we have carefully performed the mass calculation
to fourth order, we will cite only the final answer to simplify our 
presentation. 
For a more detailed discussion of all the operators which contribute the hyperon masses at this order, 
see~\cite{WalkerLoud:2004hf,Tiburzi:2004rh,Tiburzi:2005na}.  
It is straightforward to generalize the operators contained in these works to the 
relevant $SU(2)$ multiplets of hyperons.

%
\subsection{Masses of $S=1$ Baryons \label{s:S=1and2M}}

The mass of the $\L$ baryon to $\mathcal{O}(\e^4)$ is given by
\begin{align}
M_{\L} =&\ 
	M_{\L}^{(0)}
	+\frac{\s_{\L}}{(4\pi f_\pi)} \, m_\pi^2
	-\frac{ g_{\pi\L\S}^2}{(4 \pi f_\pi)^2} \cF(m_\pi, \D_{\L \S}, \mu)
	-\frac{4 g_{\pi\L \S^*}^2}{(4 \pi f_\pi)^2} \cF(m_\pi, \D_{\L \S^*}, \mu)
\nonumber\\&\ 
	+\frac{3 g_{\pi\L \S}^2 (\s_\L -  \s_\S)}{2( 4 \pi f_\pi)^3}  
		m_\pi^2  \cJ(m_\pi,  \D_{\L \S}, \mu)
	+\frac{6 g_{\pi\L \S^*}^2 (\s_\L -  \ol \s_\S)}{( 4 \pi f_\pi)^3}
		m_\pi^2  \cJ(m_\pi, \D_{\L \S^*},\mu)
\nonumber\\&\ 
	+\frac{m_\pi^4}{(4 \pi f_\pi)^3} \left[
		\a^{(4)}_{\L} \log \frac{m_\pi^2}{\mu^2}  + \b^{(4)}_{\L} \right],
		\label{eq:Lambda}
\end{align}
where the nonanalytic functions $\cF(m,\d,\mu)$ and $\cJ(m,\d,\mu)$ have been given in~\cite{Tiburzi:2005na}. 
The coefficients $\a_\L^{(4)}$ and $\b_\L^{(4)}$ are linear combinations of unknown
LECs as well as finite contributions from lower-order operators.
Additionally we have written the $\L$ mass (as well as the remainder of the hyperon masses)
in terms of the physical values of $m_\pi$ and $f_\pi$. 
As the $\cO(\e^2)$ mass spurion involves a quark mass insertion scaled by $f$, 
we have used the one-loop expression for the pion mass and pion decay constant
$f_\pi$ to arrive at Eq.~\eqref{eq:Lambda}.

The mass of the $\S$ baryon is given by
\begin{align}
M_{\S} =&\
	M_{\L}^{(0)} + \D_{\L \S} 
	+ \frac{\s_{\S}}{(4\pi f_\pi)} \, m_\pi^2
	- \frac{2  g_{\pi\S\S}^2}{(4 \pi f_\pi)^2} \pi m_\pi^3
			- \frac{ g_{\pi\L\S}^2}{3 (4 \pi f_\pi)^2} \cF(m_\pi, - \D_{\L \S}, \mu)
\notag \\&\ 
		-\frac{4 g_{\pi\S^* \S}^2}{3(4 \pi f_\pi)^2} \cF(m_\pi, \D_{\S \S^*}, \mu)
			+\frac{2 g_{\pi\S^* \S}^2 (\s_\S -  \ol \s_\S)}{( 4 \pi f_\pi)^3}
		m_\pi^2  \cJ(m_\pi, \D_{\S \S^*}, \mu)
\notag\\&\ 
	+\frac{g_{\pi\L \S}^2 (\s_\S -  \s_\L)}{2( 4 \pi f_\pi)^3}  
		m_\pi^2  \cJ(m_\pi,  - \D_{\L \S}, \mu)
					+\frac{m_\pi^4}{(4 \pi f_\pi)^3} \left[
			\a^{(4)}_{\S} \log \frac{m_\pi^2}{\mu^2}  + \b^{(4)}_{\S} \right]\, .
\end{align}
Finally the mass of the $\S^*$ baryon has the form
\begin{align}
M_{\S^*} =&\ 
	M_{\L}^{(0)} + \D_{\L \S^*}
	+ \frac{\ol \s_{\S}}{(4\pi f_\pi)} \, m_\pi^2
	- \frac{10}{9} \frac{ g_{\pi\S^*\S^*}^2}{(4 \pi f_\pi)^2} \pi m_\pi^3
	+\frac{m_\pi^4}{(4 \pi f_\pi)^3} \left[
		\a^{(4)}_{\S^*} \log \frac{m_\pi^2}{\mu^2}  + \b^{(4)}_{\S^*} \right]
\notag \\ &\ 
	- \frac{2}{3(4 \pi f_\pi)^2} \Big[ g_{\pi\S^* \S}^2 \cF(m_\pi,-\D_{\S \S^*} ,\mu)
	+ g_{\pi\L \S^*}^2 \cF(m_\pi, - \D_{\L \S^*}, \mu) \Big]
\notag\\&\ 
	+\frac{(\ol \s_\S - \s_\S) m_\pi^2 }{( 4 \pi f_\pi)^3}
		\Big[ g_{\pi\S^* \S}^2 \cJ(m_\pi, -\D_{\S \S^*}, \mu)
	+g_{\pi\L \S^*}^2 \cJ(m_\pi, - \D_{\L \S^*}, \mu) \Big]\, .
\end{align}

%
\subsection{Masses of $S=2$ Baryons \label{s:S=2M}}

Carrying out the calculation for the $\Xi$, we find its mass takes the form
\begin{align}
M_{\X} =&\  
	M_{\X}^{(0)}
	+ \frac{\s_{\X}}{(4\pi f_\pi)} \, m_\pi^2
	- \frac{3 g_{\pi\X \X}^2}{(4 \pi f_\pi)^2} \pi m_\pi^3
	- \frac{2 g_{\pi\X^* \X}^2}{(4 \pi f_\pi)^2} \cF(m_\pi, \D_{\X \X^*},\mu)
\nonumber\\&\ 
		+\frac{3 g_{\pi\X^* \X}^2 (\s_\X - \ol \s_\X)}{( 4 \pi f_\pi)^3}  m_\pi^2 \cJ (m_\pi, \D_{\X \X^*}, \mu)
	+\frac{m_\pi^4}{(4 \pi f_\pi)^3} \left[
		\a^{(4)}_\X \log \frac{m_\pi^2}{\mu^2}  + \b_\X^{(4)}  \right],
\end{align}
and the mass of the $\X^*$
has the form
\begin{align}
M_{\X^*} =&\ 
	M_{\X}^{(0)}
	+ \D_{\X \X^*}
	+ \frac{\ol \s_{\X}}{(4\pi f_\pi)} \, m_\pi^2
	- \frac{5}{3} \frac{ g_{\pi\X^*\X^*}^2}{(4 \pi f_\pi)^2} \pi m_\pi^3
	- \frac{g_{\pi\X^* \X}^2}{(4 \pi f_\pi)^2} \cF(m_\pi, -\D_{\X \X^*},\mu)
\nonumber\\&\ 
		+\frac{3 g_{\pi\X^* \X}^2 (\ol \s_\X - \s_\X)}{2( 4 \pi f_\pi)^3}  m_\pi^2 \cJ (m_\pi, - \D_{\X \X^*}, \mu)
+\frac{m_\pi^4}{(4 \pi f_\pi)^3} \left[ 
		\a^{(4)}_{\X^*} \log \frac{m_\pi^2}{\mu^2}  + \b^{(4)}_{\X^*} \right].
\end{align}

%
\subsection{The $\O^-$ Mass\label{s:S=3M}}

\begin{figure}[tb!]
\epsfig{file=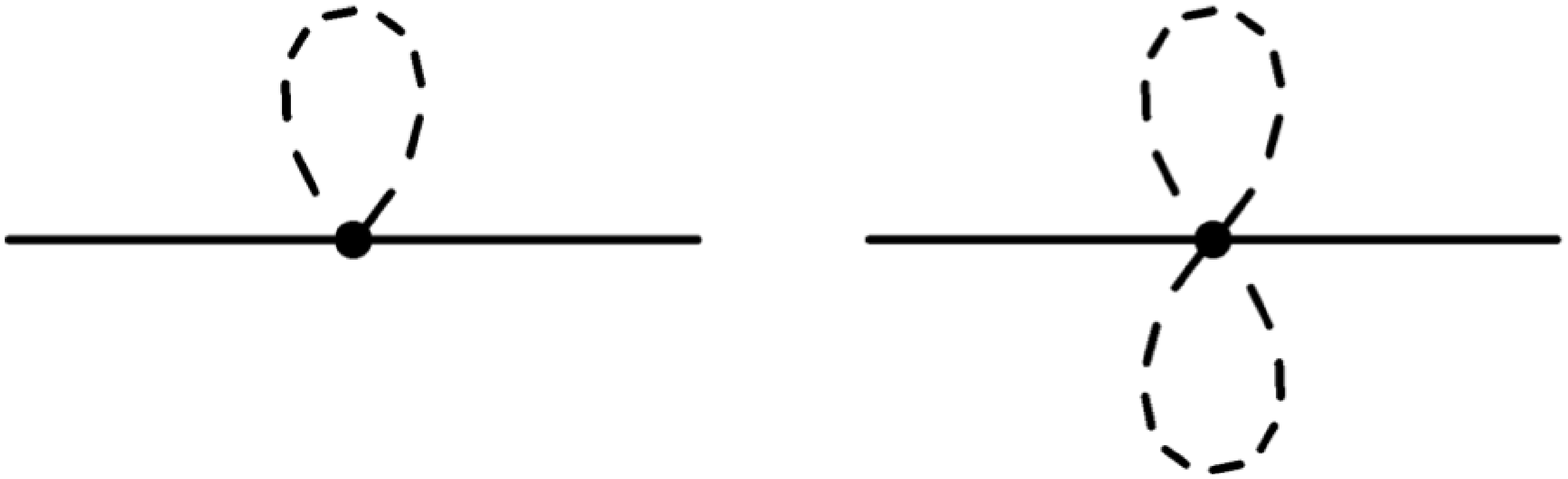,angle=270,width=6.5cm}
$\qquad \qquad$
\epsfig{file=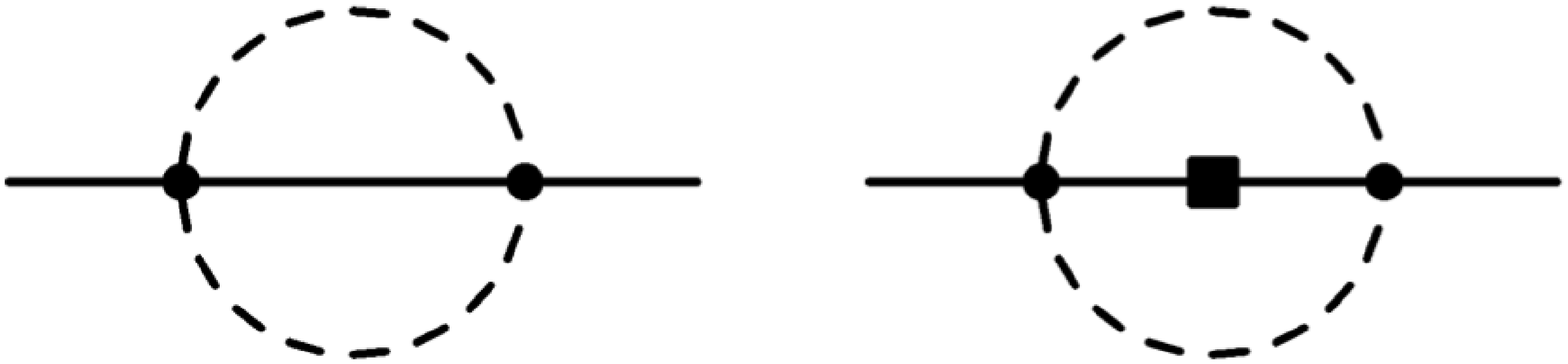,angle=270,width=6.5cm}
\caption{On the left:
loop diagrams needed up to 
$\cO(\e^6)$ 
to determine the 
$\O^-$ 
mass.  
The tadpole is generated by expanding the 
$\cO(\e^2)$ 
and 
$\cO(\e^4)$ 
terms of the Lagrangian to second order, 
while the two-headed tadpole
is generated by expanding the
$\cO(\e^2)$ operators to fourth order.  
On the right:
contributions to the $\O^-$ mass involving an internal $\O^-$. 
The circles denote insertions of the mass operator with coefficient $\ol \s_\O$. 
The first diagram depicts a contribution at $\cO(\e^7)$, 
while the relativistic correction to the $\O^-$ propagator (denoted by a square)
yields the second diagram which makes a contribution
at $\cO(\e^8)$. 
}
\label{f:omega} 
\end{figure}

Using the baryon Lagrangian
in the $S=3$ sector, we can 
easily determine 
the $\O^-$ mass to 
sixth order in the pion mass.
To work at this order, we need 
to distinguish between the 
physical pion mass dependence
which arises from renormalized loop
propagators, and quark-mass insertions
that arise from terms in the Lagrangian.
We choose to express all quantities in 
terms of physical (as opposed to bare) parameters.  
For example, the tree-level contribution at 
$\cO(\e^2)$
is a quark-mass insertion. 
To work to sixth order, we
convert the quark mass and pion decay constant 
to the physical
pion mass 
and physical decay constant 
using the known two-loop results~\cite{Burgi:1996qi,Bijnens:2006zp}.
At $\cO(\e^4)$, 
there are local contributions from Eq.~\eqref{eq:S=3L},
and a tadpole graph which must
be evaluated, see Fig.~\ref{f:omega}.
One must use one-loop expressions at this order to convert the 
bare pion mass and decay constant to their physical values.
The final contributions are at $\cO(\e^6)$.  
There are local contributions, 
tadpole contributions, 
and a double tadpole, see Fig.~\ref{f:omega}.
The latter arises from expanding the leading-order
mass operator to sixth order.

%
%
%
%
%
%
%

Assembling the results of this calculation, 
we find
\begin{eqnarray}
M_\O
&=& 
M_{\O}^{(0)}
+
\frac{m_\pi^2 }{(4\pi f_\pi)}
\a^{(2)}_\O
+
\frac{m_\pi^4}{(4 \pi f_\pi)^3} 
\left[
\a^{(4)}_\O
\log \frac{m_\pi^2}{\mu^2} 
+ 
\b^{(4)}_\O
\right]
\notag \\ && 
+
\frac{m_\pi^6}{(4 \pi f_\pi)^5}
\left[
\a^{(6)}_\O
\log^2 \frac{m_\pi^2}{\mu^2}
+ 
\b^{(6)}_\O
\log \frac{m_\pi^2}{\mu^2} 
+
\gamma^{(6)}_\O
\right].
\end{eqnarray}
There are six independent parameters
in this expression, one of which
is the $SU(2)$ chiral limit
omega mass $M_\O^{(0)}$.
The remaining five parameters are linear
combinations of the low-energy constants
in Eq.~\eqref{eq:S=3L},
with one relation among the 
$\a^{(j)}_\O$'s, specifically:
\begin{equation}
\a^{(2)}_\O
=
\ol \sigma_\O,
\quad
\a_\O^{(4)} 
=
- 6 \ol \s_\O - 3  t_\O^A,
\quad \text{and} \quad
\a_\O^{(6)}
=
28 \ol \s_\O
+ 22  t_\O^A\, .
\end{equation}

The form of the expansion of
$M_\Omega$ 
suggests that, provided the 
coefficients are of natural size,
the $\Omega^-$ 
mass will be as well-behaved
as pion observables calculated 
in $SU(2)$. 
While to the order we work, there are no 
odd powers of the pion mass contributing to 
$M_\O$, 
this is not true at to all orders. 
Indeed the first non-vanishing odd
power of the pion mass arises at 
$\cO(m_\pi^7)$.
We show an example of a diagram leading 
to such a contribution
in Fig.~\ref{f:omega}. 
The first relativistic corrections arise
at one higher order in the expansion, 
for example from the kinetic insertion
depicted in the figure.

%
\section{Phenomenology}

The expansion parameter for $SU(3)$ heavy baryon $\chi$PT is given approximately by $\e \sim 0.5$,
leading us to question the perturbative series.
Despite this reservation, 
we know phenomenological examples where  $SU(3)$ symmetry holds to a remarkable level.  
Such an example is the Gell-Mann--Okubo (GMO) relation. 
Using the charge-neutral baryon masses in the relation
\begin{equation}\label{eq:GMO}
	\D_\textrm{GMO} = M_\L + \frac{1}{3}M_{\S^0} - \frac{2}{3}M_n - \frac{2}{3}M_{\Xi^0}\, ,
\end{equation}
one observes that $\D_\textrm{GMO} = 10.3(2) \texttt{ MeV}$.  
Normalizing by the centroid octet baryon mass, 
$M_B = \frac{1}{8}M_\L +\frac{3}{8}M_{\S^0} +\frac{1}{4}M_n +\frac{1}{4}M_{\Xi^0}$, one finds
$\d_\textrm{GMO} \equiv \D_\textrm{GMO} /M_B = 0.0090(2)$,
The smallness of this effect can be understood in part with the use of the Ademollo-Gatto Theorem~\cite{Ademollo:1964sr}, a consequence of which is that any operator transforming as an \textbf{8} under $SU(3)$ will automatically satisfy the GMO relation.  This relation was recently explored with lattice QCD~\cite{Beane:2006pt}, using pion masses in the range $m_\pi \sim 293 - 592$~\texttt{MeV}.
For a fixed strange quark mass, they found
$\D_\textrm{GMO} \rightarrow 0$, 
as the pion mass is increased towards the $SU(3)$ point.

On the other hand, 
the first lattice calculation of hyperon axial charges was recently performed~\cite{Lin:2007ap},
finding deviations from $SU(3)$ symmetry.
This lattice calculation employed pion masses in the range 
$m_\pi \sim 354 - 754$~\texttt{MeV}, 
where the heaviest mass point was at the $SU(3)$ symmetric point 
$m_\pi = m_K$.  
A relation, $\d_{g_{\pi BB}}$, between axial couplings was explored%
\footnote{In comparing our axial couplings to those of~\cite{Lin:2007ap}, be careful to note 
$g_{\pi NN} = g_A$, $g_{\pi\S\S} = 2g_{\S\S}$, and $g_{\pi\Xi\Xi} = -g_{\Xi\Xi}$.} 
\begin{equation}\label{eq:deltag}
	\d_{g_{\pi BB}} = g_{\pi NN} - g_{\pi\S\S} - g_{\pi\Xi\Xi} \, ,
\end{equation}
which must vanish in the $SU(3)$ symmetric limit.  
While the study found that this quantity vanished at the $SU(3)$
point, the value extrapolated to the physical point was non-vanishing, $\d_{g_{\pi BB}} = 0.23$.
This deviation from the $SU(3)$ 
limit is more characteristic of expected $SU(3)$ breaking effects than the observed deviation from the GMO relation, 
Eq.~\eqref{eq:GMO}.

%
\subsection{Matching the $SU(2)$ and $SU(3)$ Lagrangians \label{s:phenomenology}}

One wonders whether the $SU(3)$ relations among baryon masses and axial charges are satisfied
when these quantities are determined using only the constraints imposed by $SU(2)$.  
To answer such questions, we match the $SU(2)$ Lagrangian onto the $SU(3)$ heavy baryon Lagrangian.
To $\cO(\e^2)$, this Lagrangian was originally given in~\cite{Jenkins:1990jv,Jenkins:1991es}. 
Carrying out the matching, one finds
\begin{align}
&M_N^{(0)} = M_0^{SU(3)} -(b_D - b_F + b_0) \left(2m_K^2 - m_\pi^2 \right),&
	&\frac{\s_N}{(4\pi f)} = -(b_D + b_F + 2b_0),&
\nonumber\\
&M_\L^{(0)} = M_0^{SU(3)} -(\frac{4}{3}b_D + b_0) \left(2m_K^2 - m_\pi^2 \right),&
	&\frac{\s_\L}{(4\pi f)} = -(\frac{2}{3}b_D + 2b_0),&
\nonumber\\
&M_\S^{(0)} = M_0^{SU(3)} -b_0 \left(2m_K^2 - m_\pi^2 \right),&
	&\frac{\s_\S}{(4\pi f)} = -2(b_D + b_0),&
\nonumber\\
&M_\Xi^{(0)} = M_0^{SU(3)} -(b_D +b_F +b_0) \left(2m_K^2 - m_\pi^2 \right),&
	&\frac{\s_\Xi}{(4\pi f)} = -(b_D -b_F + 2b_0).&
\end{align}
Note the $\cO(\e^2)$ relation $2Bm_s = 2m_K^2 - m_\pi^2$.  
Using these tree-level matchings, the GMO relation, Eq.~\eqref{eq:GMO} is exactly satisfied.  
This in turn implies two relations for the $M_B^{(0)}$, and $\s_B$ LECs. 
Given the $SU(2)$ LECs, $\{M_N^{(0)}, M_\L^{(0)}, M_\S^{(0)}, M_\Xi^{(0)} \}$ and $\{ \s_N, \s_\L, \s_\S, \s_\Xi \}$ determined from lattice QCD calculations, 
we can ask how well they satisfy the GMO relation.
We will return to this question below, after completing the matching in this section.

Using $SU(3)$ symmetry, one can form two relations analogous to the GMO relation for the decuplet baryons, 
namely $\D M_T$, and $\D M_T^\prime$ given by
\begin{align}
	&\D M_T \equiv 2M_{\S^*} - M_\D - M_{\Xi^*},&
	&\D M_T^\prime \equiv M_{\S^*} + M_\O - 2M_{\Xi^*}.&
\end{align}
When one matches the $SU(2)$ Lagrangian onto $SU(3)$ at tree level, one finds four
relations among the $M_T^{(0)}$, and $\ol \s_T$ LECs.  
These relations guarantee that $\D M_T=0$ and $\D M_T^\prime=0$.
Of perhaps the most phenomenological interest are the pion-baryon couplings.  
To tree level,
 these are simply given by the axial couplings in the $\mathcal{O}(\e)$ Lagrangian.  
 Matching onto $SU(3)$ we find the tree-level relations:
\begin{align}\label{eq:SU2toSU3gBB}
&g_{\pi NN} = D+F,& 
&g_{\pi\S\S} = 2F,&
&g_{\pi\Xi\Xi} = D-F,&
&g_{\pi\L\S} = 2D,&
\\ \label{eq:SU2toSU3gBT}
&g_{\pi\D N} = \mathcal{C},&
&g_{\pi\S^*\S} = \frac{1}{\sqrt{3}}\mathcal{C},&
&g_{\pi\Xi^*\Xi} = \frac{1}{\sqrt{3}}\mathcal{C},&
&g_{\pi\L\S^*} = -\frac{1}{\sqrt{2}}\mathcal{C},&
\\ \label{eq:SU2toSU3gTT}
&g_{\pi\D\D} = \mathcal{H},&
&g_{\pi\S^*\S^*} = \frac{2}{3}\mathcal{H},&
&g_{\pi\Xi^*\Xi^*} = \frac{1}{3}\mathcal{H}.&
\end{align}

%
\subsection{The pion mass dependence of the hyperon masses at $\cO(\e^3)$ \label{s:MB_NLO}}

With numerical estimates of the axial couplings, 
we can ascertain the nonanalytic $\cO(\e^3)$ 
contributions to the hyperon masses for pion masses currently accessible with lattice QCD.  
These size of such contributions, moreover, all us to assess the behavior of the perturbative expansion.
Experimentally, only two of the spin--1/2 hyperon axial charges are known.%
\footnote{
Our notation for axial charges might be potentially confusing. 
We have denoted the axial charges by $g_{\pi BB}$, which are only 
perturbatively close to the pion-baryon-baryon couplings.
}  
Equating the experimental values to our chiral limit parameters, 
we have:
$g_{\pi NN} = 1.27$, and $g_{\pi\L\S} = 1.47$.
The $\S$ and $\Xi$ axial charges are not experimentally known.  
We can, however, use the values predicted from the recent lattice calculation~\cite{Lin:2007ap}.  
Adjusting for our differing conventions, and setting the lattice extrapolated values
equal to our chiral limit parameters, we have: $g_{\pi\S\S} = 0.78$, and $g_{\pi\Xi\Xi} = 0.24$.

The values of the axial-transition couplings can be estimated from comparing 
the widths derived from the mass formulae given in Sec.~\ref{s:S=1and2M},
and the experimentally known widths in the PDG~\cite{Yao:2006px}.
We determine:
$g_{\pi\D N} = 1.48$, 
$g_{\pi\Xi^*\Xi} = 0.69$,
$g_{\pi\L\S^*} = -0.91$,
and
$g_{\pi\S^*\S} = 0.76$.
Comparing these values with the expectations from $SU(3)$, we observe from ratios 
$(g_{\pi\D N} -\sqrt{3}g_{\pi\Xi^*\Xi} )\, /\, ( g_{\pi\D N} +\sqrt{3}g_{\pi\Xi^*\Xi} ) = 0.11$,  \emph{etc.},  that 
$SU(3)$ violations are on the order of 5-10\%.
These violations are
noticeably smaller than those observed for the octet baryon axial charges, Eq.~\eqref{eq:deltag}.  
We have neither a phenomenological nor lattice determination of the axial charges of the $\S^*$ and $\Xi^*$ resonances.%
\footnote{
It was recently stressed that even the $SU(2)$ axial coupling
$g_{\pi \D \D}$ is poorly known~\cite{Jiang:2008we}. 
We hope future lattice calculations will remedy this. 
}
We use the estimate of $\cH=2.2$ from~\cite{Butler:1992pn}, 
and impose the tree-level $SU(3)$ relations 
to determine $g_{\pi \S^* \S^*}$ and $g_{\pi \X^* \X^*}$.

\begin{figure}
\begin{tabular}{cc}
\includegraphics[width=0.35\textwidth]{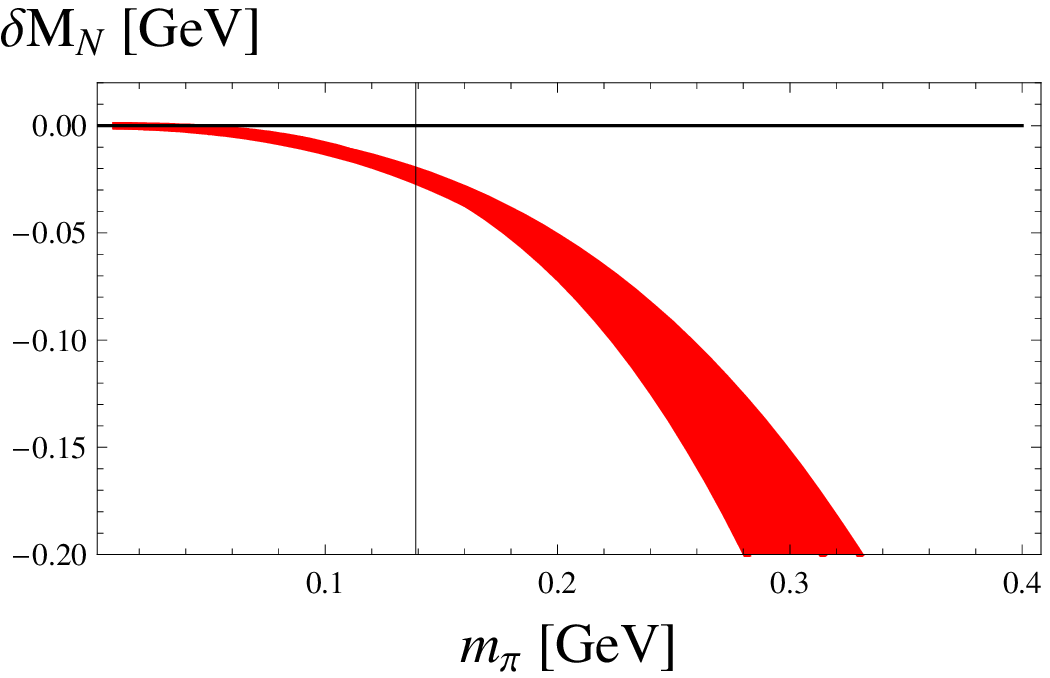}
&\includegraphics[width=0.35\textwidth]{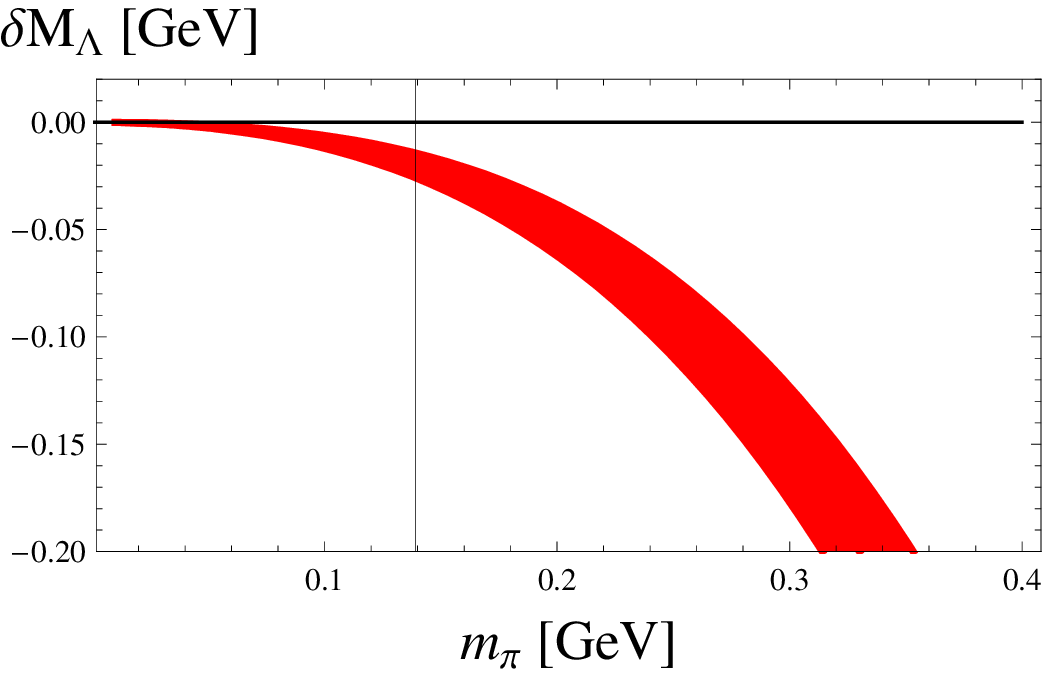}
\\
$(a)$ & $(b)$ \\
\includegraphics[width=0.35\textwidth]{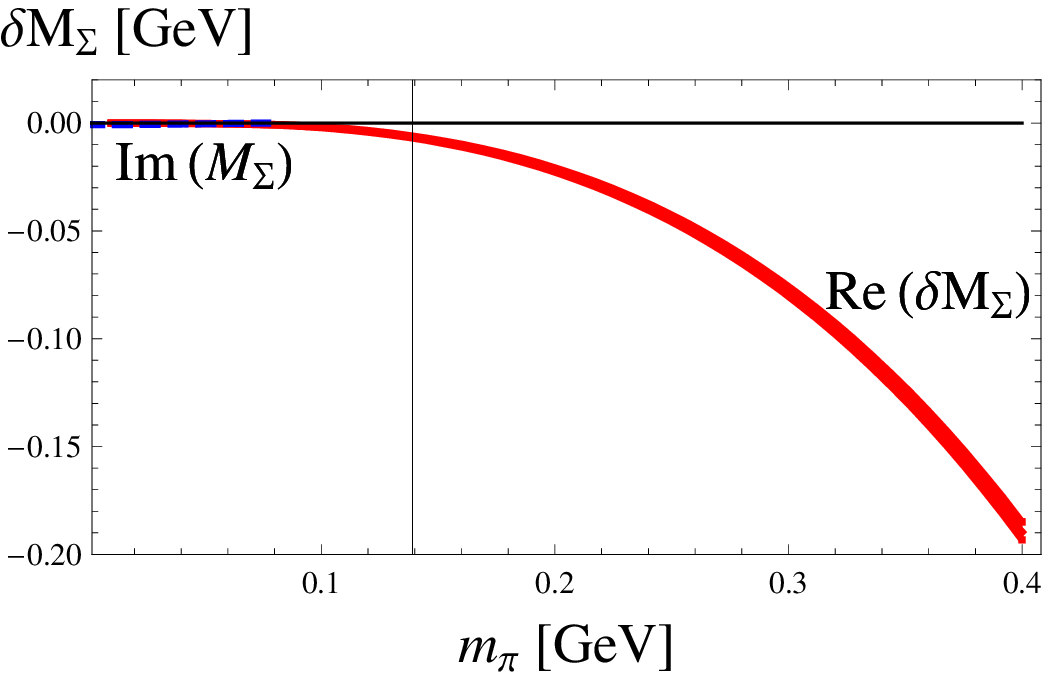}
&\includegraphics[width=0.35\textwidth]{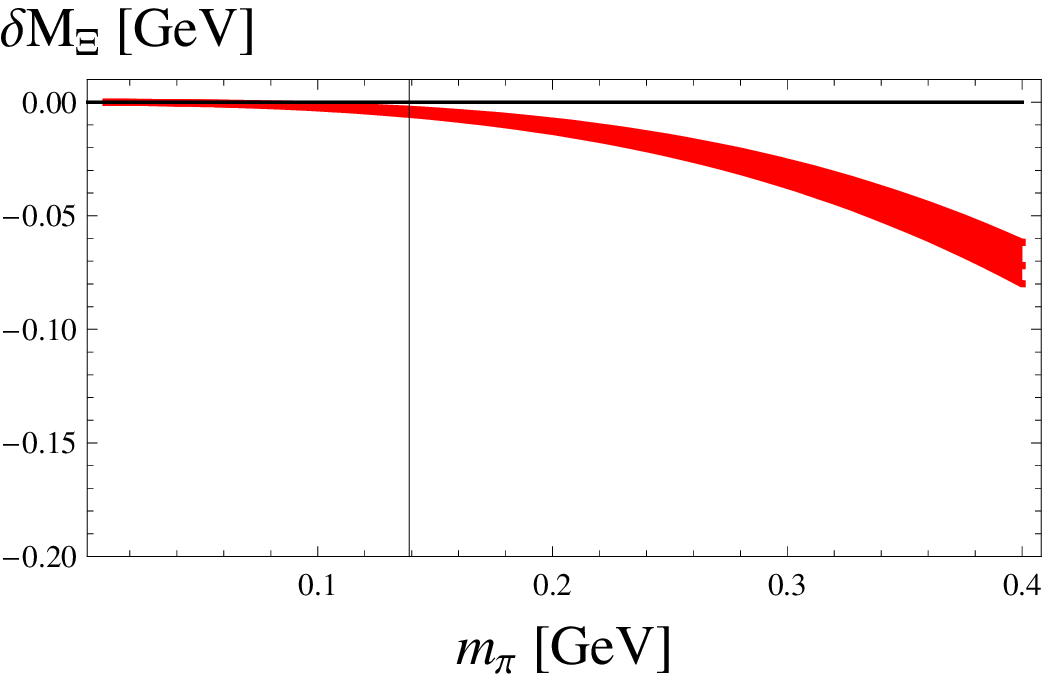}
\\
$(c)$ & $(d)$ 
\end{tabular}
\begin{tabular}{ccc}
\includegraphics[width=0.32\textwidth]{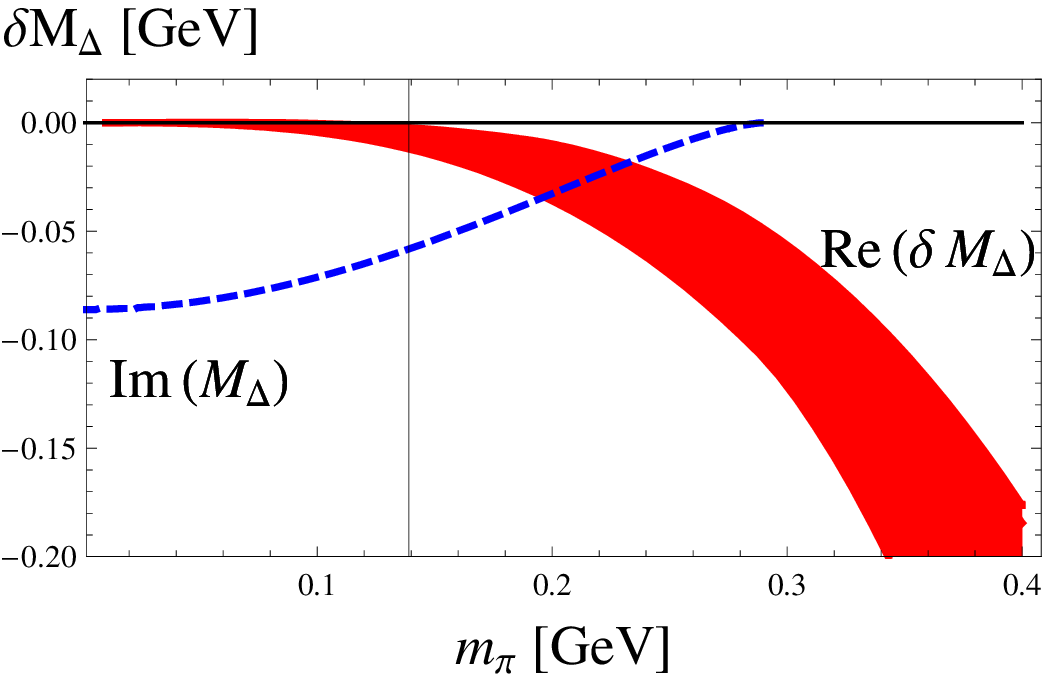}
&\includegraphics[width=0.32\textwidth]{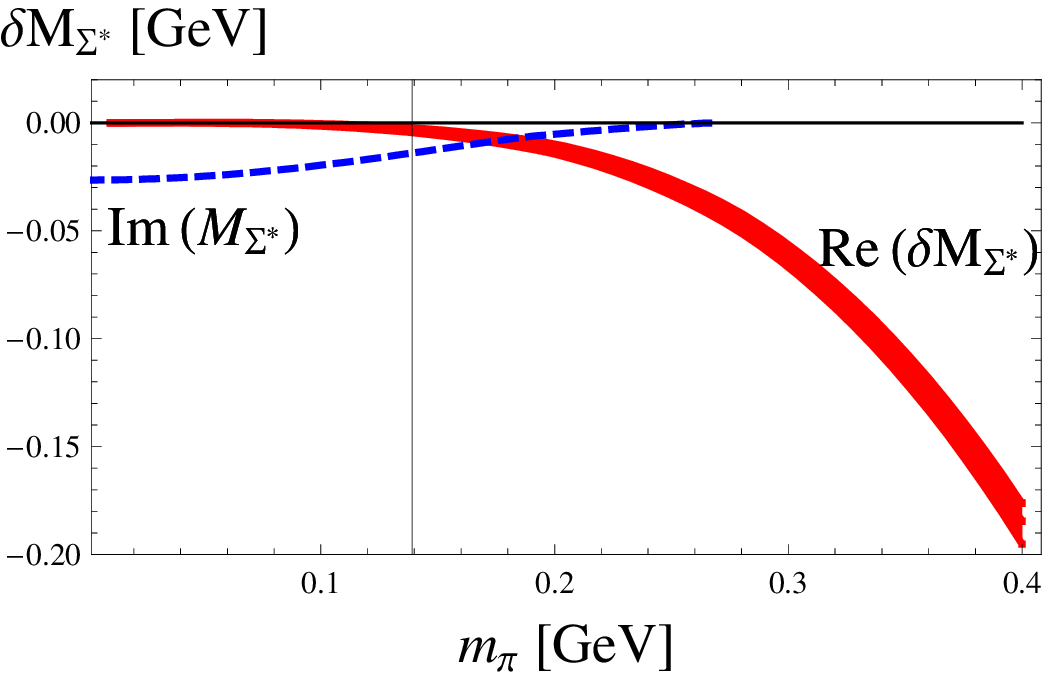}
&\includegraphics[width=0.32\textwidth]{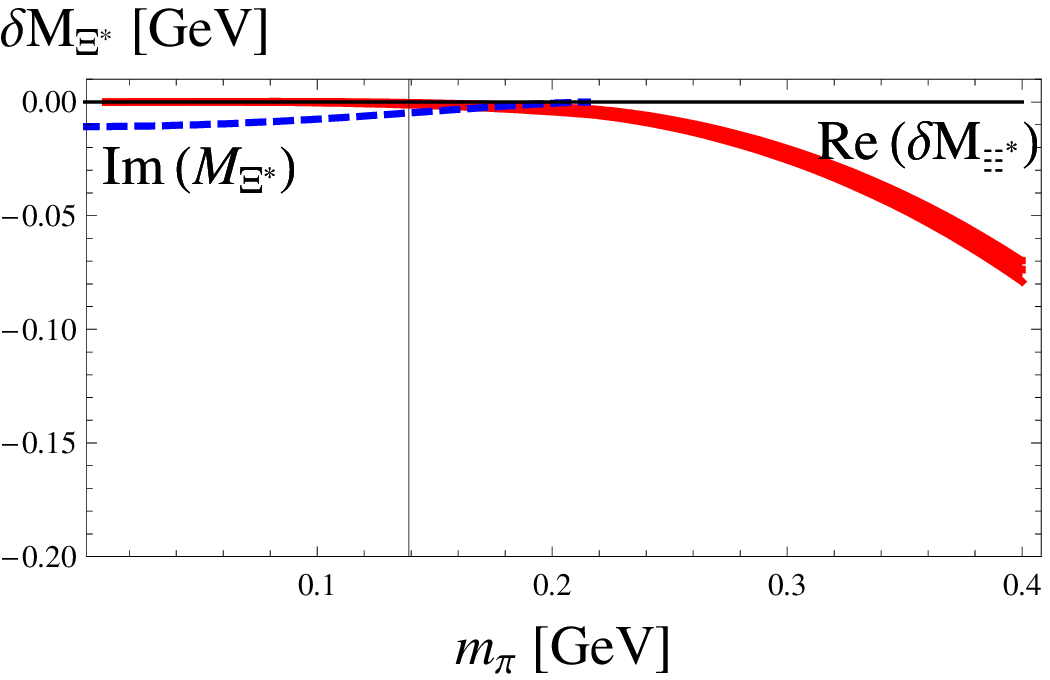}
\\
$(e)$ & $(f)$ & $(g)$
\end{tabular}
\caption{\label{f:NLOHyperonMasses}
Pion mass dependence of nonanalytic mass corrections. 
In (a)--(d), we plot the third order contribution to the masses
of the $N$, $\L$, $\S$, and $\Xi$ baryons, respectively.
The third-order mass contribution to the spin-3/2 resonances, 
the $\D$, $\S^*$, and $\Xi^*$ baryons, are plotted in (e)--(g). 
Uncertainty from the unknown local contribution at third order
is shown by the band, and is generated by varying the renormalization
scale from $\mu = 0.75$ to $1.25 \, \texttt{GeV}$.
Expressions for the $N$ and $\D$ masses in $SU(2)$ can be found 
in~\cite{Tiburzi:2005na}, for example.
}
\end{figure}

With this phenomenological and lattice input, we can estimate the third order chiral corrections to the masses of the $\L$, $\S$ and $\Xi$ baryons.  This is done in Fig.~\ref{f:NLOHyperonMasses}, where the nucleon is also included for comparison purposes.  We do not have complete information at third order because there are unknown local contributions.
%
An easy way to estimate the size of such coefficients is to vary the renormalization scale $\mu$. 
In the figure, we plot the pion mass dependence of the third order correction. We let $\mu$ vary between 
$0.75$ and $1.25 \, \texttt{GeV}$.  Additionally plotted is the imaginary part of the $\S$ mass for a pion mass $m_\pi < \D_{\L\S}$.  This contribution is $\mu$-independent.  In the case of the $\S$, the imaginary part is quite small as the
available phase space is restrictive, even in the chiral limit.
At the physical pion mass, chiral corrections appear to be very perturbative.  We see that these $\cO(\e^3)$ corrections are 
mildest for the $\Xi$ baryon mass.  One expects the $SU(2)$ theory will converge better with increasing strangeness, 
although the $\L$ mass does not appear to behave much better than the nucleon.  
The $\cO(\e^3)$ correction to the mass of the $\Xi$, on the other hand, is $\lesssim 5\%$  
for pion masses up to $m_\pi \sim 400$~\texttt{MeV}.

We can perform a similar analysis for the decuplet baryon masses.  
We have also plotted the mass corrections at $\cO(\e^3)$ 
for the $\D$, $\S^*$, and $\Xi^*$ in Fig.~\ref{f:NLOHyperonMasses},
including the imaginary parts as well. 
As with the octet baryon masses, 
we observe an improved convergnece in the decuplet baryon masses with increasing
strangeness.
In fact, as discussed in Sec.~\ref{s:S=3M}, the $\cO(\e^3)$ corrections to the $\O$ mass are identically zero.

%
\subsection{The Gell-Mann--Okubo relation and lattice QCD \label{s:GMO}}

We can go further, and attempt to extract
the chiral limit masses and leading-order sigma terms from available lattice data. 
The use of $SU(3)$ $\chi$PT to analyze lattice hyperon masses is quite limited, 
with the first explorations performed in Ref.~\cite{Frink:2005ru}.  
In Ref.~\cite{Beane:2006pt}, excellent agreement between the lattice GMO relation 
and the predictions from $SU(3)$ heavy baryon $\chi$PT were found.  
Most recently, convergence issues were found in an NLO $SU(3)$ analysis of the LHPC hyperon 
specturm~\cite{WalkerLoud:2008bp}.  
The recently published results of  LHPC have lattice results for pion masses ranging from 
$0.295$ to $0.757 \, \texttt{GeV}$~\cite{WalkerLoud:2008bp}. 
For our purposes, we use their lightest four pion masses to fit to the second-order expressions for
the spin-$1/2$ hyperon masses.  
We assign a somewhat arbitrary 20\% systematic error to account for expected extrapolation errors, 
most importantly from the truncation of the chiral expansion to such a low order.  We find
\begin{eqnarray}
M_N^{(0)} &=& 1.02(1)(20) \, \texttt{GeV},  \qquad \s_N = 0.141(4)(28),
\notag\\
M_\L^{(0)} &=& 1.18(1)(24) \, \texttt{GeV}, \qquad \s_\L = 0.099(6)(20),
\notag\\
M_\S^{(0)} &=& 1.28(1)(26) \, \texttt{GeV}, \qquad \s_\S = 0.068(5)(14),
\notag\\
M_\Xi^{(0)} &=& 1.38(1)(28) \, \texttt{GeV}, \qquad \s_\Xi = 0.045(3)(9)\, .
\end{eqnarray}
Putting these into the GMO relation, one finds
\begin{align}
&M_\L^{(0)} +\frac{1}{3}M_\S^{(0)}-\frac{2}{3}M_N^{(0)} - \frac{2}{3}M_{\Xi}^{(0)} = 6 \, \texttt{MeV},& &\text{and}&
&\s_\L +\frac{1}{3}\s_\S -\frac{2}{3}\s_N -\frac{2}{3}\s_{\Xi} &=& -0.002.&
\end{align}
The resulting values of these LECs are in remarkable agreement with the GMO formula, even when the $SU(3)$ symmetry is not imposed.%
\footnote{As discussed in~\cite{WalkerLoud:2008bp}, the $\cO(\e^2)$ formula are not sufficient to describe the chiral extrapolation of the hyperon masses to the physical point.  There are expected sizable changes to these LECs when higher-order terms are included in the extrapolation, and it is expected one will need to perform the extrapolation to at least $\mathcal{O}(\e^4)$.
There are not enough lattice results with light pion masses to reliably perform such extrapolations, so we do not undertake this endeavor here.}

\section{Summary \label{s:summy}}
Above we have developed two-flavor 
\CPT\ 
to describe hyperons. 
The hyperons are embedded into
$SU(2)$ 
multiplets and treated 
in the non-relativistic
heavy baryon expansion. 
Terms in 
$SU(3)$  
\CPT\
that scale as 
$m_\eta / M_B$, 
or 
$m_K / M_B$ 
have effectively been summed to all orders
in the two-flavor theory.
The resulting 
$SU(2)$ 
\CPT\ is expected to behave better.
In particular, the heavy baryon approximation
has an expansion parameter that scales as 
$m_\pi / M_B$.
This expansion parameter decreases
with increasing strangeness, and we 
expect that the 
$SU(2)$
theory of hyperons will
converge better than
the $SU(2)$ 
theory of nucleons. 
A striking example is that of the 
$\O^-$.
Up to sixth order, there are no odd powers
of the pion mass in the chiral expansion of
$M_\O$. 
We can thus anticipate that the usefulness of the 
two-flavor expansion for the 
$\O^-$
will be competitive with the 
two-flavor expansion of pion properties. 

While the two-flavor theories of hyperons we developed
offer a marked improvement over the 
$SU(3)$ 
theory of baryons, 
lattice data are required to extract values of the 
LECs. 
Given current phenomenological and lattice knowledge, 
the remaining couplings to be determined in the 
$S=1$ and $S=2$ sectors are the
leading sigma terms (the $\s$ or $\ol \s$) for each 
multiplet, and values of the spin-$3/2$ resonance
axial charges, $g_{\pi\S^*\S^*}$ and $g_{\pi\X^* \X^*}$. 
LECs for the $\O^-$ are almost completely unknown. 
Knowledge 
of these constants will help extrapolate
the $\O^-$ mass to the physical pion mass;
and, in turn, tune the strange quark 
mass to its physical value, or alternatively use $M_\O$ to set the scale in lattice calculations.

Using the theory we have devoloped, other hyperon observables can be calculated in the $SU(2)$ chiral expansion, 
for example the hyperon axial charges~\cite{OrginosEtal}.
Additionally, the strangeness changing sector needs revisiting, in particular the old puzzle of non-leptonic hyperon decays.  
Furthermore for lattice QCD applications, the $SU(4|2)$ partially quenched~\cite{Chen:2001yi,Beane:2002vq} and mixed action~\cite{Bar:2002nr,Bar:2003mh,Tiburzi:2005vy,Bar:2005tu,Tiburzi:2005is,Chen:2006wf,Chen:2007ug} Lagrangians should be constructed.  The relevant baryon multiplets are simply
generalizations of those containing one or two heavy quarks~\cite{Arndt:2003vx,Tiburzi:2004kd,Mehen:2006vv}.
Finally, 
comparing with lattice data and, when possible, experiment, will help determine whether $SU(3)$ relations emerge in the two flavor theory, even if the $SU(3)$ expansion is ill-fated.

\begin{acknowledgments}
We thank Paulo Bedaque for spiritual guidance.  
This work is supported in part by the 
U.S.~Dept.~of Energy,
Grants 
No.~DE-FG02-93ER-40762
(B.C.T. and A.W.-L.),
and
No.~DE-FG02-07ER-41527
(A.W.-L.).
\end{acknowledgments}

\bibliography{hb,su3}


\end{document}